\documentclass[11pt]{article}
\usepackage{erice,epsfig}

\def\be{\begin{equation}}
\def\ee{\end{equation}}
\def\bea{\begin{eqnarray}}
\def\eea{\end{eqnarray}}

\begin{document}
\vspace*{4cm} \title{HOT TOPICS IN ULTRA-PERIPHERAL ION COLLISIONS}

\author{G. BAUR$^a$, C. A. BERTULANI$^b$, 
M. CHIU$^c$, I. F. GINZBURG$^d$, K. HENCKEN$^e$, S. R. KLEIN$^f$,\break
J. NYSTRAND$^g$,K. PIOTRZKOWSKI$^h$
C. G. ROLDAO$^i$ D. SILVERMYR$^g$, J. H. THOMAS$^f$,\break
S. N. WHITE$^j$ and P. YEPES$^k$}

\address{
$^a$ Forschungzentrum J\"ulich, J\"ulich, Germany\break
$^b$ Natl. Superconducting Cyclotron Lab., Michigan State University,
East Lansing, MI, 48824 USA \break
$^c$ Columbia University, New York, NY, 10027 USA \break
$^d$ Sobolev Institute of Mathematics, SB RAN, 630090, Novosibirsk, 
Russia\break
$^e$ Universit\"at Basel, 4056 Basel, Switzerland\break
$^f$ Lawrence Berkeley National Laboratory, Berkeley, CA, 94720, USA\break
$^g$ Department of Physics, Lund University, Lund SE-22100, Sweden\break
$^h$ Departement de Physique, Universite Catholique de Louvain, 
B-1348 Louvain-la-Neuve, Belgium \break
$^i$ Instituto de Fisica Teorica, Universidade Estadual Paulista,
Sao Paolo, Brazil \break
$^j$ Brookhaven National Laboratory, Upton, NY, 11973, USA \break
$^k$ Physics and Astronomy Dept., Rice University, Houston, TX, 77005, USA
}
\vskip .1 in

\maketitle\abstracts{Ultra-peripheral collisions of relativistic heavy
ions involve long-ranged electromagnetic interactions at impact
parameters too large for hadronic interactions to occur.  The nuclear
charges are large; with the coherent enhancement, the cross sections
are also large.  Many types of photonuclear and purely electromagnetic
interactions are possible. We present here an introduction to
ultra-peripheral collisions, and present four of the most compelling
physics topics.  This note developed from a discussion at a workshop
on ``Electromagnetic Probes of Fundamental Physics,'' in Erice, Italy,
Oct. 16-21, 2001.}

\section{What are Ultra-Peripheral Collisions?}

Ultra-peripheral collisions are interactions that occur at impact
parameters, $b$ large enough that no hadronic interactions can occur.
In simple terms, $b>2R_A$, where $R_A$ is the nuclear radius.  Only
electromagnetic interactions are possible; they can be purely
electromagnetic (`two-photon') or photonuclear($\gamma$A).

Ultra-peripheral collisions of heavy ions are interesting because the
large ion charge ($Z$) gives rise to extremely intense short-duration
electromagnetic fields.  Following the Weizs\"acker-Williams method,
these fields are usually described as a spectrum of almost real
(quasireal) photons.  When the photon wavelength is larger than the
nucleus, the emission is coherent and the flux is proportional to
$Z^2$.  For relativistic particles with Lorentz boosts $\gamma$ (to
the lab frame), this occurs for lab-frame photon energies $k <
\gamma\hbar c/R_A$. In the target nucleus rest frame, photons from the
other nucleus have a maximum energy of $(2\gamma^2-1)\hbar c/R_A$,
about 500 GeV at the Relativistic Heavy Ion Collider (RHIC) at BNL,
and 1 PeV at the Large Hadron Collider (LHC) at CERN.  RHIC and LHC
are high-luminosity $\gamma A$ colliders; LHC has an energy reach far
beyond other existing or planned machines.  Figure 1 shows the ratio
of the $\gamma A$ to $AA$ luminosity for heavy ions at RHIC and the
LHC; $\gamma p$ collisions may also be of interest.  Also shown, for
comparison, is the ratio for the proposed eRHIC electron-ion collider.

The maximum $\gamma\gamma$ collision energy $W_{\gamma\gamma}$ is
$2\hbar c\gamma/R_A$, about 6 GeV at RHIC and 200 GeV at the LHC.
Figure 2 compares the $\gamma\gamma$ flux at the LHC with the
completed LEP II $e^+e^-$ collider at CERN.  The LHC will have a
significant energy and luminosity reach beyond LEPII, and could be a
bridge to $\gamma\gamma$ collisions at a future $e^+e^-$ linear
collider.

These cross section enhancements extend the physics reach of heavy ion
colliders.  For example, the coherent $\rho$ production rate at RHIC
is $120 s^{-1}$ with gold, rising to $230,000 s^{-1}$ with calcium at
the LHC.  The LHC is a vector meson factory.  The cross section for
$e^+e^-$ production reaches 200 kb for lead at the
LHC.~\cite{reviews} Many processes are unique to heavy ion colliders,
either because of the high$-Z$ beam particles or because of
interference between $\gamma\gamma$ and coherent $\gamma A$ processes.

Ultra-peripheral processes have striking signatures.  Two-photon and
photon-Pomeron processes lead to final states with a small number of
centrally produced particles, with rapidity gaps (regions of phase
space containing no particles) separating the central final state from
both beams.  In addition, neither nucleus should be broken up.  For
coherent photon emission, the photon perpendicular momentum $p_T$ is
less than $\hbar/R_A$.  If the coupling to both nuclei is coherent,
the transverse momentum $p_T$ of the final state is less than
$2\hbar/R_A\approx 100$ MeV/c.  Even incoherent interactions (with
respect to one of the nuclei) such as photoproduction of heavy quarks
will be characterized by a single rapidity gap and intact nucleus.

We present four compelling physics topics for ultra-peripheral
collisions: gluon shadowing in nuclei, Pomeron coupling to nuclei,
interferometry with short-lived particles and searches for new
physics, most notably by studying the $\gamma WW$ vertex.  Numerous
other interesting subjects, many unique to heavy ion collisions, are
not included, either because they require specialized apparatus to
study, or because the physics is not yet clearly defined.  For
example, electromagnetic production of $e^+e^-$ pairs tests QED in the
non-perturbative, strong-field regime created by the high-$Z$ sources,
but requires a dedicated experiment.  Production of lepton or hadron
pairs accompanied by capture of the negatively charged particle also
fall into this category. Interference between identical
final states (like $\pi^+\pi^-$ or $e^+e^-$) produced via
$\gamma\gamma$ and $\gamma A$ channels is sensitive to the relative
phases of many processes.  This interference is very interesting, but
we don't know exactly what to measure yet. Additional details and
other peripheral collisions physics are presented in several recent
reviews.~\cite{reviews}

\begin{figure}
\center{\psfig{figure=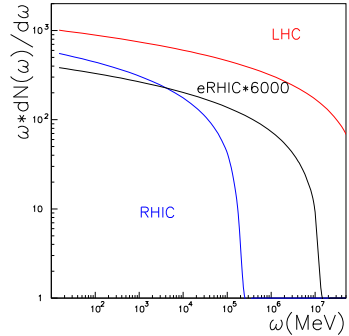,height=1.8in,clip=}}
\caption[]{The relative $\gamma A$:$AA$ luminosity at RHIC and the LHC,
compared with that expected at the proposed eRHIC.~\cite{eic} The
eRHIC curve has been multiplied by 6000, the ratio of the planned
eRHIC luminosity to the RHIC design luminosity.  The photon energy
$\omega$ is that in the target nucleus rest frame.}
\label{fig:gammaa}
\end{figure}

\section{What is the most interesting physics?}

These four topics presented are highlighted for their strong physics
interest and their feasibility with existing and planned detectors
(possibly with slight modifications).  

\subsection{Gluon Shadowing in Nuclei}

The parton distributions (structure functions) of heavy nuclei is of
interest because the low-$x$ gluon density is high, and, in fact,
gluon saturation should be present.  These strong fields could be a
colored glass condensate, with new and unexpected
properties.~\cite{glass} Even in the absence of a condensate, accurate
measurements of the parton distributions are important in
understanding the initial states of relativistic heavy ion collisions.

To date, most measurements of parton shadowing in nuclei have relied
on lepton deep inelastic scattering (DIS).  DIS is only sensitive to
the quark content of the nuclei; the gluon content is inferred from
the $Q^2$ evolution of the structure functions.  Furthermore, the only data
on nuclear structure functions is from fixed target experiments,
limited to fractional momenta $x>10^{-3}$ and low momentum transfer.

The gluon distributions can be measured at heavy ion colliders by
studying photoproduction of heavy (charm and
bottom;~\cite{photoprod,bbar} top is also observable at the
LHC~\cite{top}) quarks; this reaction occurs primarily by photon-gluon
fusion.  RHIC will be able to extend the previous measurements to
higher $Q^2$, while the LHC will be able to reach much smaller $x$ and
higher $Q^2$.  The LHC measurements will probe regions where many
models predict that gluon saturation will be observed.  For example,
in a colored glass model,~\cite{colorglass} photoproduction of charm
and bottom at the LHC is much less than is expected in conventional
parton distributions.

In photoproduction, the photons are almost real, but the high mass and
$p_T$ of the produced heavy quarks introduces a high $Q^2$ scale.
These events can be separated from heavy quark production in grazing
hadronic interactions by requiring one rapidity gap in the event,
along with one undisturbed nucleus, with a charged particle
multiplicity cut to eliminate more central hadronic collisions.

The heavy quark production cross sections are sensitive to the heavy
quark mass and to higher-order corrections.  Despite this, RHIC and
the LHC can offer a `first look' at as yet unexplored regions of $x$
and $Q^2$.  More importantly, by comparing the heavy quark production
in $AA$ (effectively $\gamma A$) and $pA$ (effectively $\gamma p$)
collisions, the shadowing of nuclear particle distributions can be
measured.  In the ratio, the QCD uncertainties cancel, and the
achievable accuracy should be sufficient for a meaningful measurement
of shadowing. It is difficult to obtain these structure functions from
other methods. The leading alternative, studies of jets, direct
photons and the like in $pA$ collisions at RHIC and the LHC suffers
from large systematic uncertainties.

\subsection{Pomeron Couplings to Nuclei}

\begin{figure}
\center{\psfig{figure=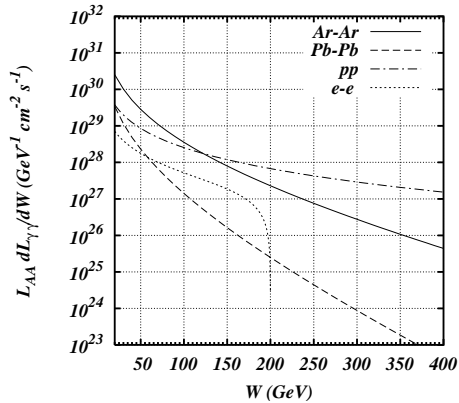,height=2.2in,clip=}}
\caption[]{The effective $\gamma\gamma$ luminosity as a function of
the $\gamma\gamma$ energy $W_{\gamma\gamma}$ at the LHC, compared with
that at LEP II.  The $AA$ and $e^+e^-$ luminosities are $ArAr$:
$5.2\times10^{29} cm^{-2}s^{-1}$, $PbPb$: $4.2\times10^{26}
cm^{-2}s^{-1}$, $pp$: $10^{33} cm^{-2}s^{-1}$ (corresponding to about
1 collision/ beam crossing; the LHC $pp$ design luminosity is 10 times
higher), and $e^+e^-$: $5\times10^{31} cm^{-2}s^{-1}$ at an $e^+e^-$
center of mass energy of 200 GeV.}
\label{fig:gammagamma}
\end{figure}

Many hadronic interactions do not involve color exchange.  Examples
include proton-proton elastic scattering, photoproduction of vector
mesons, and other diffractive phenomena, including jet and $W^\pm$
production.  These processes are generally described in terms of
Pomeron exchange.

The Pomeron has a long and complex history.  It is generally agreed
that it involves the strong force, but is colorless, and that it has
the quantum numbers of the vacuum ($J^{PC} = 0^{++}$).  Beyond that,
there is a great diversity of views as to its
structure.~\cite{niccolo} There are two classes of Pomeron models
corresponding to hard and soft interactions. The soft Pomeron
represents the absorptive part of the scattering cross section.
Often, this is dressed up in Regge theory.  It describes much data,
including elastic scattering and photoproduction of vector mesons, but
lacks internal detail, and a connection to QCD.  In photoproduction of
vector mesons, a photon fluctuates to a $q\overline q$ pair.  This
pair then scatters elastically (via Pomeron exchange) from the
nucleus, emerging as a vector meson.

The hard Pomeron model usually applies for higher momentum transfers,
where QCD is expected to be reasonably perturbative.  The hard Pomeron
is usually modeled as two gluons, with additional structure, such as
a gluon ladder.

One test of these models is their coupling to nuclei.  The soft
Pomeron couples equally to all nucleons.  For heavy states like
$J/\psi$ and $\Upsilon$, the cross section for coherent photonuclear
production of vector mesons scales as $A^2$. For lighter mesons, a
single $q\overline q$ state can interact with multiple Pomerons; this
can be calculated accurately in the Glauber model.  In the
`black-disk' limit (very large $q\overline q$ states), the cross
section scales as $A^{4/3}$.  

However, if the Pomeron is made of gluons, the situation changes.  The
gluon distributions of nucleons change when they are placed in nuclei
(the EMC effect/shadowing).  If Pomerons are made of gluons, and if
the gluon content of nuclei is different from that in independent
nucleons, then the Pomeron coupling should change.  These couplings can
be studied systematically by comparing production rates of different
mesons at different $p_T$ from different targets.  For $p_T <
\hbar/R_A$, the mesons couple coherently to the entire nucleus, while
at higher $p_T$, the coupling is to individual nucleons.  The two
types of interactions have different properties, and will provide
useful cross-checks.

Measurements at LHC will extend the photoproduction cross
section measurements to energies significantly above the HERA regime.
HERA data for the $J/\psi$ finds $\sigma\approx
W^{0.8}$.~\cite{jpsi} This rapid increase with $\gamma p$
center-of-mass collision energy $W$ cannot continue indefinitely or
the $J/\psi$ cross section will exceed the total photoproduction cross
section.  A similar rise is seen for the $\Upsilon$, albeit with lower
statistics.~\cite{upsilon} The $\gamma p$ cross sections can be
measured with $pA$ collisions; LHC can reach an order of magnitude
beyond HERA in $W$.  

Heavy meson production in $AA$ ($\gamma A$) collisions is of interest
because of its sensitivity to gluon saturation.~\cite{frankfurt} The
LHC can study nuclear vector meson photoproduction at previously
unavailable energies.  The cross section is a probe of the gluon
density in very low-$x$, high-$A$ region where many models predict
that saturation should be observed.  The meson mass and $p_T$
variation will cover the ranges usually described by the hard and soft
Pomerons.

\subsection{Interferometry with Short-Lived Particles}

For coherent $\gamma A$ processes, the two nuclei act as an
interferometer, with two nuclei serving as sources.  Because the
electromagnetic field has a long range, and the hadronic scattering a
short range, vector meson production is well localized at the two
sources, especially compared with the typical impact parameters of
20-60 fm at RHIC, rising to 20-250 fm at the LHC. The sign of the
interference depends on the final state parity.

Because most of the vector mesons have lifetimes far shorter than the
time needed to travel between the two nuclei, amplitudes from the two
sources must decay independently.  This seems to preclude
interference; consider, for example, one source decaying
$J/\psi\rightarrow e^+e^-$ and the other $J/\psi\rightarrow
\pi^+\pi^-\pi^0$. Yet, quantum mechanics requires that they interfere.
The resolution to this paradox is that the post-decay wave functions
must contain amplitudes for all possible decay modes: branching
ratios, angular distributions, and even decay times. Because this
requires a spatially distributed (non-factorizable) wave function, the
system is an example of the Einstein-Podolsky-Rosen paradox.  It is a
test of quantum mechanics for unstable particles.  The interference is
destructive for $\vec{p}_T \cdot \vec{b}< \hbar$, so the vector meson
production is reduced for $p_T < \hbar/\langle
b\rangle$.~\cite{interf} As the data from the STAR detector in Fig. 3
shows, in this region, a clean sample of mesons with $p_T < \hbar/R_A$
can be selected with simple cuts.~\cite{parkcity}

\begin{figure}
\center{\psfig{figure=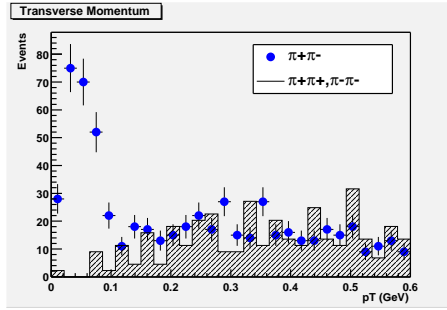,height=1.6in,clip=}}
\caption{The $p_T$ spectrum of two-track events observed
by the STAR detector.  The large peak for $p_T<100$ MeV/c
is expected for coherent production of $\rho$ mesons.}
\label{fig:mbpt}
\end{figure}

Multiple vector meson production is also observable in the strong
fields.  The probability of $\rho$ production in grazing collisions
$\mu$ is about $1\%$ with gold at RHIC, rising to 3\% with lead at the
LHC.~\cite{vmprod} If each meson is treated independently, then the
probability of producing $n$ $\rho$ at RHIC and LHC are $P(n) =
\exp{(-\mu)}\mu^n/n!$, or, for $n=2$, $(1\%)^2/2$ and $(3\%)^2/2$, for
$n=3$, $(1\%)^3/6$ and $(3\%)^3/6$. The total rates are obtained by
integrating $P(b)$ over impact parameter; the doublet rates are large
($10^6$/year for $\rho^0\rho^0$ at both RHIC and the LHC), while the
triplets are still observable.

These samples can be used to study quantum correlations; the final
state is considerably richer than for single mesons.  The two source
ions can each emit single or multiple photons.  Production
correlations should produce Hanbury-Brown Twiss (HBT)-like
enhancements at small relative momentum.  However, because the vector
mesons can be produced literally on top of each other, many new
phenomena may be observable.  For example, the vector mesons could
interact with each other, leading to new final states such as
$\rho^+\rho^-$. Finally, since for relative momenta, $\delta p <
\hbar/\tau$, coherence will be maintained until the mesons decay,
correlated (stimulated) decays will be possible.

Vector meson lasing occurs for $\mu\ge1$, depending on the source
geometry.  Unfortunately, $\mu=1$ seems unreachable.  However, even
superradiant behavior should be visible even at the expected
$\mu=0.01-0.03$. As the previous paragraph shows, this leads to very
interesting phenomenology.

The strong field sources and unique geometry (two identical meson
sources) produce a new configuration - an interferometer for
short-lived particles.  The system is new, and much theoretical and
experimental work is required to exploit it.  Experimentally, clean
samples of coherent vector meson production have been extracted, and
initial measurements of interference are not far off.

\subsection{Searches for New Physics}

Two-photon physics at the LHC allows for a wide range of searches for
new physics; as Fig. 2 shows, the LHC will have an energy/luminosity
reach considerably beyond LEP2.  The triple gauge coupling $\gamma WW$
can be studied using the copious two-photon W pair production.  This
is believed to be the most sensitive process to the new physics of
large extra dimensions.~\cite{newp} Two-photon production of $Z$ pairs
or the Higgs boson is less copious, but could complement more
conventional studies at the LHC, and test for deviations from the
Standard Model predictions. In addition, two photon production rates
of the supersymmetric charginos and sleptons are
significant.~\cite{snep}

Because of the high center of mass energies, luminosity and running
time, searches for new physics with $\gamma\gamma$ interactions will
be most competitive with $pp$ mode at the LHC (see Fig. 2).  The
hadronic backgrounds can be reduced by requiring that there be
rapidity gaps in the event and tagging the forward scattered
protons.~\cite{pp} The LHC magnets will act as magnetic spectrometers
for the scattered protons, allowing the reconstruction of the photon
momenta.  This will give an unbiased estimate of the photon-photon
collision energy.  It will also allow for significant background
rejection.

Tagging will also detect double-Pomeron interactions.  Two-photon
interactions can be separated from these double-Pomeron interactions
because the Pomerons have a larger average transverse momentum than
the photons.  Depending on the final state under consideration, either
single or double tagging may be used.  The anticipated luminosity of
the tagged two-photon collisions at $W_{\gamma\gamma} > 100$ GeV
reaches then almost 1\% of the $pp$ luminosity, even for rather
conservative assumptions on the tagging efficiency.~\cite{snep}

Especially at RHIC, tagged double Pomeron interactions are of strong
interest to study meson spectroscopy. For reasons that are poorly
understood, when the transverse momentum transfer from the protons are
in the same direction, conventional $q\overline q$ mesons are
produced, while when the transverse momentum transfers are in opposite
directions, many suspected exotic (non $q\overline q$) mesons are
preferentially produced.~\cite{mesons}

Single tagging is also possible for $pA$ collisions.  The
$W_{\gamma\gamma}$ reach is lower, but up $W_{\gamma\gamma}\approx$
100 GeV the $pA$ two-photon luminosity is very competitive with the
$pp$ case. This channel can also be used to cross-check two-photon
studies using $pp$ and $AA$.

\section{Conclusions}

We have presented four compelling examples of the physics accessible
in ultra-peripheral collisions. They address a variety of new physics,
including relatively straightforward, but important measurements of
nuclear shadowing, studies of the nature of the Pomeron, and tests of
quantum mechanics and searches for new physics.  Many other
hadronic measurements are possible with existing detectors.

If experiments requiring new hardware were considered, many other
processes would compete for spaces on the list.  Many interesting
atomic physics experiments are possible.  For example, near-threshold
single and multiple $e^+e^-$ pair production probes QED in a
nonperturbative regime.  Pair production with $e^-$ capture is
important because it limits the maximum luminosity achievable with
heavy ions at the LHC.

\end{document}